\documentclass[]{aiaa-tc}
\usepackage{booktabs,amsmath,amsthm,amssymb,amsbsy,latexsym,graphicx,color}
\usepackage{rotating}
\usepackage{float}
\usepackage{subfigure}
\usepackage{upgreek}
\usepackage{bm}
\usepackage{mathdots}
\usepackage{relsize}
\usepackage{caption}
\usepackage{newfloat}
\usepackage{fixltx2e}
\usepackage{nomencl}
\usepackage{multicol}
\usepackage{indentfirst}
\usepackage{etoolbox}
\usepackage{tocvsec2}
\usepackage{gensymb}
\usepackage[titletoc]{appendix}
\usepackage{appendix}
\usepackage[bbgreekl]{mathbbol}
\usepackage{graphicx}

\title{Phase Transformation Characteristics of High-Temperature Shape Memory Alloy under Tension, Compression, and Bending Actuation Cycling}

 \author
  {
  	Daniel Martin
  	\thanks{Graduate Research Assistant, Aerospace Engineering, College Station, TX 77843, USA, Student member.}\thanksibid{1}\\
  {\normalsize\itshape
   Texas A\&M University, College Station, TX 77843, USA}\\
  \and
  Lei Xu 
  \thanks{Graduate Research Assistant, Aerospace Engineering, College Station, TX 77843, USA, Student member.}\thanksibid{2}\\
  {\normalsize\itshape
	 Texas A\&M University, College Station, TX 77843, USA}\\
   \and
   Dimitris Lagoudas
   \thanks{ Distinguished Professor,Aerospace Engineering,College Station, TX 77843, USA, Faculty member.}\thanksibid{1}\\
  {\normalsize\itshape
  Texas A\&M University, College Station, TX 77843, USA}\\
  }

\begin{document}

\maketitle

\begin{abstract}
Shape Memory Alloys (SMAs) are a unique class of intermetallic alloys that can cyclically sustain large deformations and recover a designed geometry through a solid-to-solid phase transformation. SMAs provide favorable actuation energy density properties, making them suitable for engineering applications requiring a significant, repeated, work output. To facilitate the development and validation of an SMA constitutive model considering the evolving anisotropic material response for High-Temperature SMA (HTSMA), uniaxial and pure bending actuation cycling tests on HTSMA specimens are performed by a custom-built testing frames. The phase transformation characteristics for Ni$_{50.3}$TiHf$_{20}$ HTSMA under uniaxial tension/compression and four-point bending actuation cycles are investigated. The experimental results show that the polycrystalline HTSMAs has a strong tension-compression asymmetry under uniaxial actuation cycling loading conditions. Furthermore, the four-point beam bending test shows that there is an intrinsic phenomenon when HTSMAs are subjected to cyclic actuation bending conditions, i.e., the zero-strain neutral axis shifts as a result of the asymmetric tension-compression phase transformations and the asymmetric generation of TRIP strains on different sides of the beam. The conducted experiments provide invaluable information to develop and improve the SMA constitutive model considering tension-compression asymmetry and TRIP strain generation within a unified modeling effort. As future work, additional experiments on other HTSMA components, such as torque tubes and specimens with notches or cutouts, under actuation cycling would provide more comprehensive validation data and component performance for HTSMA-based actuators.

\end{abstract}

\section{INTRODUCTION}\label{sec:intro}

\subsection{Shape Memory Alloys and Applications} 

Shape Memory Alloys (SMAs) exhibit a reversible solid-to-solid phase transformation between the austenite and martensite phases, facilitated through inputs of stress, temperature, or a thermomechanical combination. Although SMAs have a lower actuation frequency compared to other active materials, they are useful in actuation applications due to their large energy density, and their ability to recover extremely large transformation strains. In other words, SMA actuators can undergo recoverable shape changes under high stress levels and produce a large work output compared to other solid state actuator materials, such as piezoelectric ceramics or shape memory polymers.



\begin{figure}[h!]
	\centering
	\includegraphics[width=0.65\textwidth]{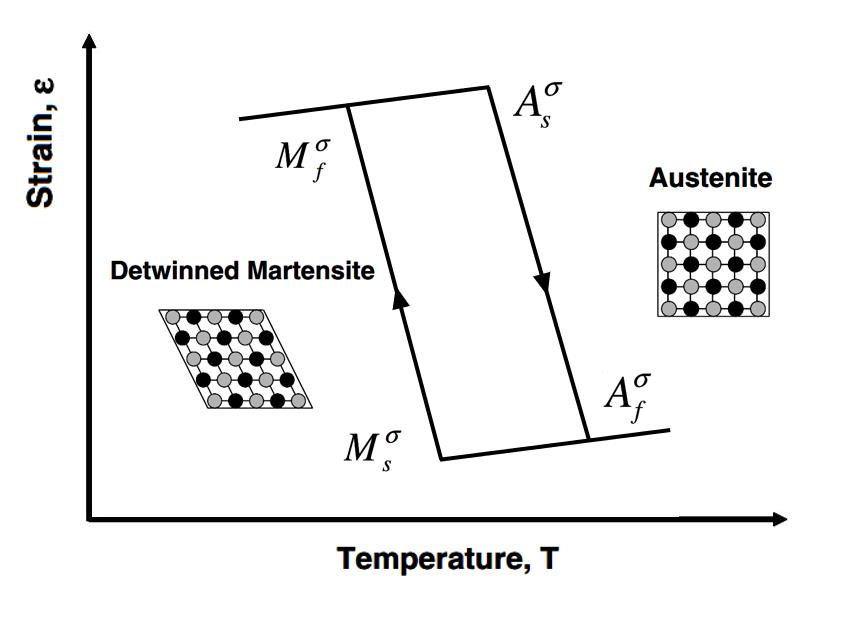}
	\caption{A representative actuation response of SMAs at a constant stress level \cite{lagoudas2008}.}
	\label{fig1:Actuation}
\end{figure}

In general, SMAs can be characterized by a phase diagram based on their phase transformation between the austenite and martensite. The four critical transformation temperatures, $M_f$, $M_s$, $A_s$, $A_f$, describe the beginning and final transition of the SMA between the two phases. Depending on the material composition, heat treatment, and processing, the transformation temperatures can be modified to be below freezing, or even well above 85$^{\circ}$C. The martensite phase is further broken down into twinned martensite and detwinned martensite. For twinned martensite, the mechanical load applied to the SMA is not high enough to see a macroscopic deformation, only a lattice structure rearrangement. On the other hand, the detwinned martensite phase shows a significant deformation compared to the austenite phase, between 1\% to 10\% macroscopic strains. The thermomechanically induced phase transformation of an SMA under a constant load is shown in Figure \ref{fig1:Actuation}. It is a recoverable deformation controlled by simple temperature inputs, and it allows for the SMA to actuate between the detwinned martensite phase and the austenite phase. This work focused on the macroscopic response of SMAs using this thermomechanical actuation in tension, compression, and bending.

\begin{figure}[ht!]
	\centering
	\includegraphics[width=0.9\textwidth]{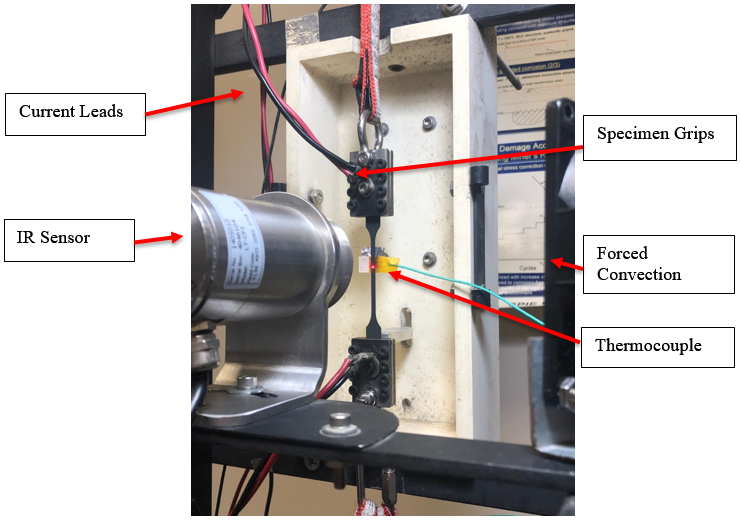}
	\caption{Test setup for actuation cycling under tension.}
	\label{fig:load_frame}
\end{figure}

SMAs are implemented in numerous commercial and industrial fields including: automotive, aerospace, electrical, robotics, and biomedical, and they can take the form of motors, actuators, transducers, structural materials, or sensors\cite{jani2014review}. The biomedical field holds the largest amount of patents and functional uses of SMAs thus far, but as the depth of understanding and manufacturing costs decrease, aerospace applications are growing, even within the restrictions of the Federal Aviation Administration. Since aircraft and spacecraft require many complex systems, the multifunctionality of SMA components have the potential to reduce complexity, volume, weight, and sometimes cost, compared to traditional electromechanical or hydraulic actuator systems \cite{hartl2007aerospace}. Some examples of SMA actuator applications for the aerospace industry are variable geometry chevrons, deployable solar panels, and active space radiators \cite{icardi2009preliminary,mabe2005design,mabe2006boeing,hartl2006thermomechanical,wheeler2015design,bertagne2018coupled}. Specifically, Sofla et al.\cite{sofla2010shape} have completed a comprehensive review of wing morphing technologies using antagonistic SMA-actuated flexural structural forms to actively bend and twist a wing for improved aerodynamic performance. The SMA actuators are capable of enduring the aerodynamic loads, satisfy power requirements, and achieve the force and torque required for a small unmanned aircraft. They show that SMAs can be incorporated in advanced systems without weight penalties or stiffness loss compared to conventional actuators. The Boeing variable geometry chevron utilizes bending actuation of SMAs in order to utilize a trade-off between noise mitigation at take-off and landing and engine performance at high altitude for a 787 engine. Instead of a static chevron, the alternative design uses passively controlled SMA actuators to adjust engine properties with changes in altitude. At low altitudes, the chevron is in a noise reduction configuration, and at high altitudes, the chevron relaxes to recover engine performance \cite{mabe2005design,mabe2006boeing,hartl2006thermomechanical}. The reconfigurable solar array uses an SMA torque tube to deploy and retract a microsat solar array system. The project showcases an SMA’s ability to reduce the weight, volume, and complexity of the system using SMAs instead of the conventional system using electronic actuators. It also satisfies control requirements where it could be deployed at low temperatures under 60 seconds with a total twist angle of 46 degrees. A prototype was built to prove the technology and the theory \cite{wheeler2015design}. SMAs are practical for space applications as well. Space radiators are important for thermal control in crewed spacecraft design, but they require high turn down ratios needed for a large range on heat rejection rates. A novel radiator concept has been designed, analyzed, and fabricated using SMA material for passive geometric reconfiguration by Bertagne et al \cite{bertagne2018coupled}. The radiator design allows for a 12:1 turndown ratio, compared to a 3:1 ratio used by current space radiators. SMAs could be used as sheets, strips, or wire forms to close and open the panel. At high temperatures, the radiation panel morphs into the maximum heat rejection shape, when temperatures decrease, the panel returns to a closed configuration, and the actuation process is fully reversible.


\subsection{High Temperature Shape Memory Alloys} 

While NiTi is the most studied SMA due to its large recoverable strain and corrosion resistance, HTSMAs such as the tertiary metal alloys NiTiHf and NiTiZr are becoming the present research focus due to their application in higher temperature ranges, mostly for aerospace applications. These alloys have transformation temperatures well above 85$^\circ$C, where the current Radio Technical Commission for Aeronautics guidelines require components to be dormant in operation environments \cite{sherer1947radio}. HTSMAs are also required to exhibit acceptable actuation strain recovery, long term stability, resistance to plastic deformation and functional fatigue, and corrosion resistance in order to be viable commercially\cite{ma2010high}. 
The NiTiHf alloy systems with more than 50 (at. \%) nickel content shows enhanced mechanical and thermal stability as a result of the formation of precipitates compared to nickel lean NiTiHf. Depending on material purity, however, these alloys can suffer from large thermal hysteresis, low strength, unstable shape memory behavior, and development of large residual strains from thermal cycling \cite{kockar2006method}. This material does not show other transformations in the martensite transition such as rhombohedral (R-phase) seen in nickel rich SMAs \cite{wheeler2016actuator}, bainite\cite{tadaki1988shape}, or the rubber-like behavior\cite{ren1997origin}.

\begin{figure}[h!]
	\centering
	\includegraphics[width=0.9\textwidth]{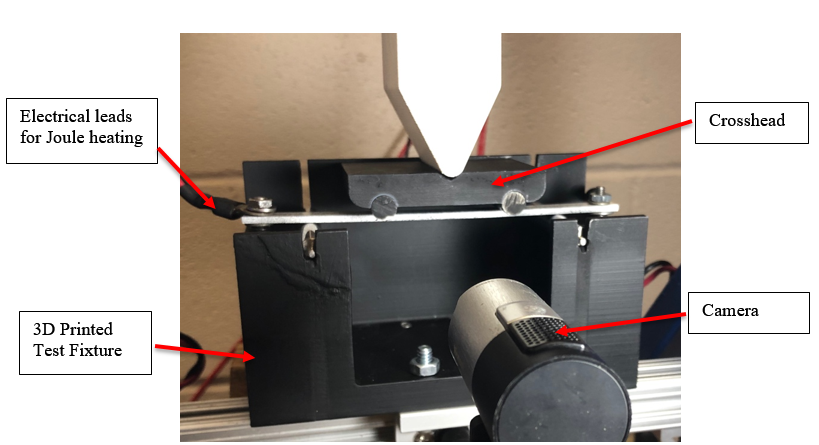}
	\caption{The custom-built four-point beam bending test setup.}
	\label{fig:4point_loadframe}
\end{figure}

\subsection{Transformation Induced Plasticity} 
After cycling from the austenite phase to the detwinned martensite phases under a thermomechanical loading path, unrecovered strains develop in SMA materials after each cycle. Unlike plastic strains in conventional metals, these remnant strains occur at significantly lower stress levels, well below the yield limit, and lead to functional fatigue of the SMA actuator. These remnant strains are due to the combination of transformation-induced plasticity (TRIP) and retained martensite. TRIP is defined as the plastic flow arising from solid state phase transformations involving shape changes at low stress levels. It is an evolutionary phenomenon during thermomechanical cycling of SMAs that leaves a number of permanent microstructural changes during the forward transformation into martensite\cite{lagoudas2008}. It is extremely prominent in the early stages of the fatigue life of the SMA or HTSMA actuator, but the development of unrecovered strain slows once dislocations reach the grain boundaries and the martensite variants stabilize. Also, TRIP is dependent on material processing and testing procedures, such as heat treatment or upper cycle temperature. It has been seen in experiments and implemented in 3-D macroscopic constitutive models\cite{Xu2017trip,xu2019AIAA}, but the characterization process must include the determination of cyclic evolution parameters in order to capture this phenomenon accurately, and this is one of the focus of this work.

\subsection{ Tension–Compression Asymmetry} 
Previous experiments have shown that polycrystalline and single crystal SMAs have a significantly different macroscopic response in compression compared to tension. Compressive loading states result in smaller displacements, higher transformation stress levels, and strain hardening \cite{gall1999tension,gall1999role}. This phenomenon has been mostly attributed to the influence of the resolved shear stress state on martensite twinning and macroscopic modes of deformation. Since the transformation initiates from a critical resolved shear stress, in the same direction of the prescribed transformation direction, there is a unidirectional dependence to the phase transformation. This creates a considerable orientation dependence and a different critical transformation stress depending on the loading direction for polycrystal SMAs with a strong crystallographic texture\cite{gall2001mechanical}. The tension-compression asymmetry has not been studied in great detail for SMAs under thermomechanical actuation, where the martensitic phase transformation is induced by temperature under a constant load. Experiments are needed to quantify the macroscopic asymmetry. Furthermore, for SMAs under bending, the compressive stress states across the thickness shows tension-compression asymmetry in the same way. The proposed work will address the relationship between the tension-compression asymmetry in uniaxial tension/compression case and in pure bending case.

\begin{table}[h] 
	\caption{Differential scanning calorimetry test matrix.}\label{tab1}\vspace{-0.5cm}
	\begin{center}
		\begin{tabular}{c|c} \toprule
			Heat treatment type & Temperate changing rate                \\           \midrule
			As received      & 10 $^{\circ}$C/min          \\
			As received + Solutionnized 950 $^{\circ}$C/ 1 hr  &10 $^{\circ}$C/min   \\
			As received + 550 $^{\circ}$C/ 3 hrs  & 10 $^{\circ}$C/min               \\
			As received + Solutionnized 950 $^{\circ}$C/ 1 hr + 550 $^{\circ}$C/ 3 hrs &  10 $^{\circ}$C/min \\
			\bottomrule
		\end{tabular}
	\end{center}
\end{table}

\begin{table}[ht!] 
	\caption{Uniaxial test matrix for dogbone specimen under tension actuation cycling.}\label{tab2}\vspace{-0.5cm}
	\begin{center}
		\begin{tabular}{c|c|c|c|c|c} \toprule
			Specimen &  Thickness, mm  & Width, mm  & Length, mm    & Stress, MPa  & Cycles          \\           \midrule
			1 &0.62&2.65&48.36&50 & 100 \\
			2 &0.85&2.84&46.57&100& 100 \\
			3 &0.76&2.59&46.16&150& 100 \\
			4 &0.77&2.68&47.75&200& 100 \\
			5 &0.85&2.73&48.45&300& 100 \\
			\bottomrule
		\end{tabular}
	\end{center}
\end{table}

\begin{table}[ht!] 
	\caption{Uniaxial test matrix for cylindrical bar specimen under compression actuation cycling.}\label{tab3}\vspace{-0.5cm}
	\begin{center}
		\begin{tabular}{c|c|c} \toprule
			Specimen &  Cross section area, mm$^2$  & Stress, MPa      \\           \midrule
			6 &11.44& 100    \\
			7 &11.40&  200   \\
			8 &11.42&  300    \\
			\bottomrule
		\end{tabular}
	\end{center}
\end{table}
 
\section{Experimental Design}\label{Model}
\subsection{Uniaxial Tension Actuation Testing}
The uniaxial actuation testing is conducted on a custom-built load control frame that allows for continued cycling of flat SMA dogbones with different heat treatment, see the test matrix in Table \ref{tab2}. The load frame, shown in Figure \ref{fig:load_frame}, can provide a constant load condition using a hanging dead load and controlled temperature cycling using resistive heating with an alternating current. The sensor data acquisition and temperature control are operated by an in-house LabView program. The diagnostics seen in the image below include an IR sensor and a contact thermocouple for temperature measurements, as well as the LVDT and DIC (not shown in the image) for displacement measurements. The load is connected directly to the bottom of the specimen grips to prevent any torsion. Computer fans are attached to the frame to provide continuous forced convection. The power system consists of a variable alternating current power supply (VARIAC) with a 130V manually tunable range, a transformer, a control box with a solid-state relay, and two pairs of electrical leads attached to the specimen grips. The grips can be tightened manually, but it is important to use a guide to prevent any twisting or bending in the dogbone when attaching the electrical leads or applying the dead load. This system supplies current between 10 and 40 A to the actuator at 0.5 to 4 V.  

\begin{figure}[ht!]
	\centering
	\includegraphics[width=0.9\textwidth]{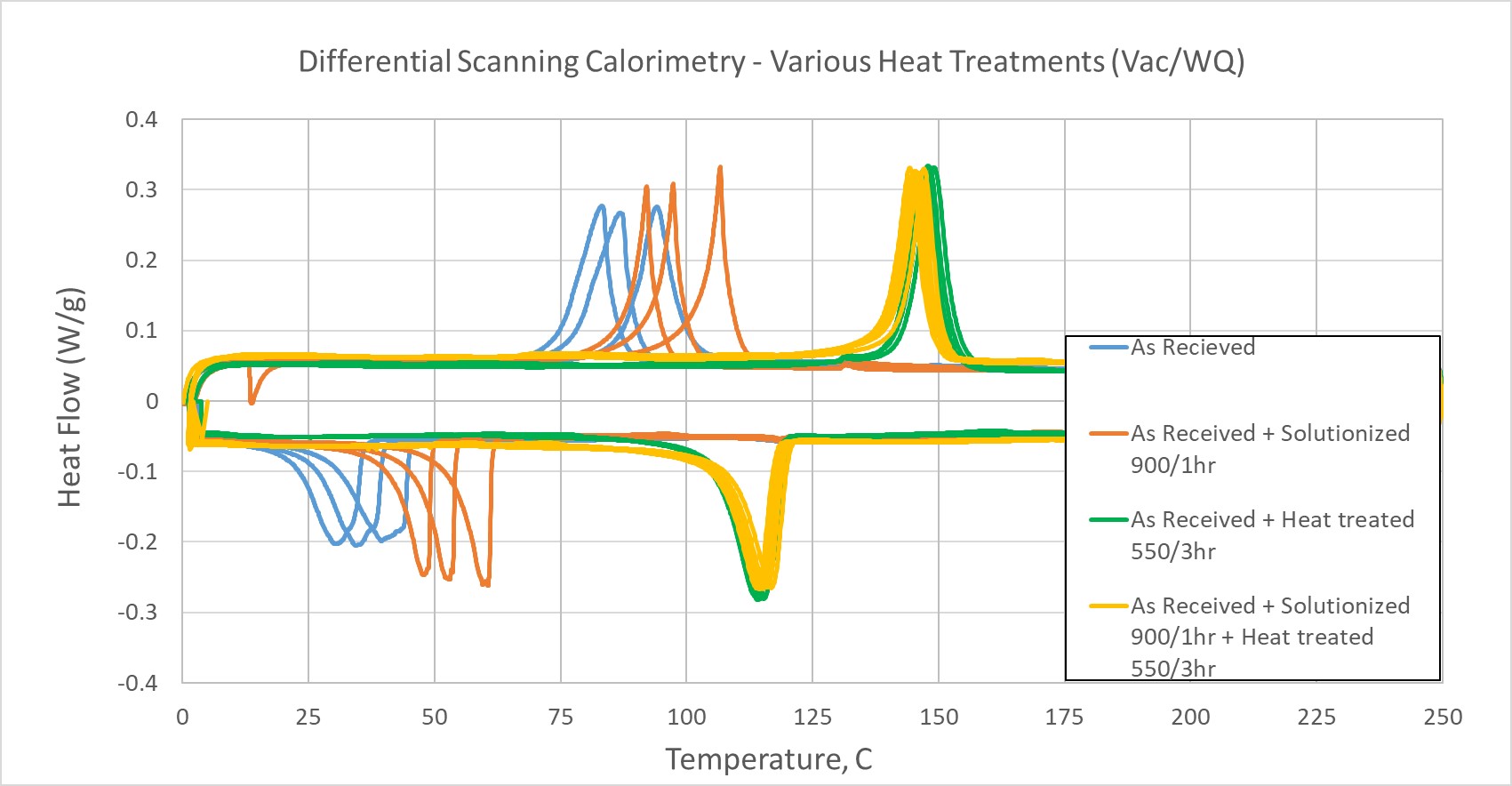}
	\caption{Differential scanning calorimetry on different heat-treated HTSMA materials.}
	\label{fig1:DSC}
\end{figure}

\subsection{Uniaxial Compression Actuation Testing}

Compression tests are conducted on cylindrical bar specimen via a MTS Insight machine using a compression plate fixture. There are two modes of control available on the insight machine: load control and displacement control. Displacement control is the desired method with the MTS Insight machine and the most accurate. For this control, a crosshead displacement rate of 0.1 mm/min is used for quasi-static testing. For actuation tests, displacement control is used to reach the desired fixed load, and then load control is used to maintain the load with a PID controller. The parameters are tweaked manually until the load cell produced a maximum error of 15.8N. The cycling control is much slower to avoid temperature gradients in the specimen, test fixture, and thermal chamber. Temperatures are extracted using three thermocouples to determine the magnitude of the temperature gradients. 


\subsection{Four-Point Bending Testing }

The next test setup is specialized for four-point bending tests. This frame, see Figure \ref{fig:4point_loadframe}, is custom-built and modified to provide cyclic actuation testing under constant loads. In order to customize for the large transformation deflection and as a cost-effective alternative to traditional bending fixtures, the test fixture is made using VeroWhite 3D printed material with stainless steel contact points and painted to allow for background contrast for DIC imaging. The stainless steel contact points do not rotate during the test in order to provide symmetry and minimized out of plane motion. The diameter of the supports and load points allow for the specimen to bend consistently during actuation cycles. The fixture, as well as the load point, is leveled and balanced. This frame could also be used for three-point bending, but this research focuses on four-point bending in order to have a pure bending loading path and to simplify the model validation procedure. In order to have similar testing procedures, the four-point bending test uses the same power supply as the uniaxial actuation tension tests. The power system consists of a variable alternating current power supply (VARIAC) with a 130V manually tunable range, a transformer, a control box with a solid-state relay, and two pairs of electrical leads attached directly to the specimen. After applying the alignment load, 1.85 N, to the beam, actuation tests show that the leads are not influencing or loading the specimen as long as they have free range of motion throughout the experiment.


\begin{figure}[ht!]%
	\centering
	\subfigure[]{%
		\includegraphics[width=0.5\textwidth]{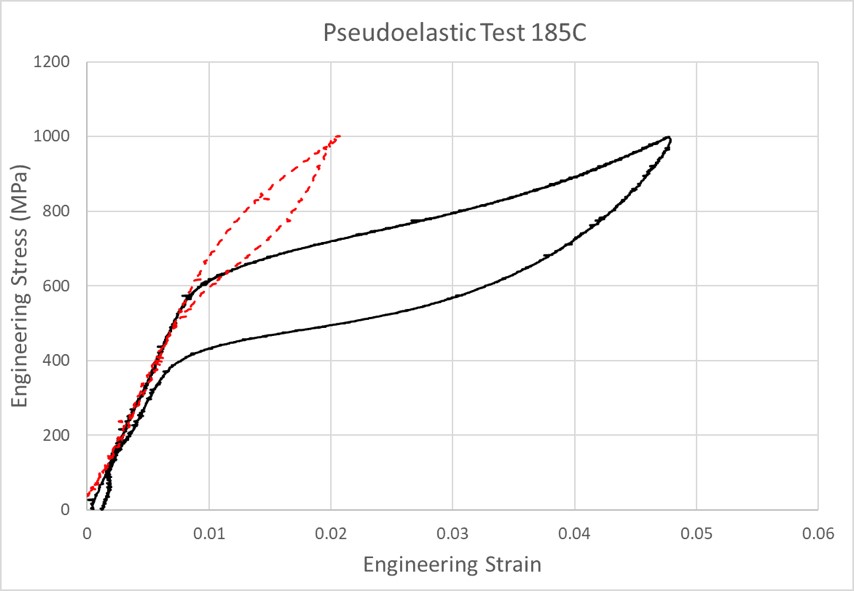}}%
	\hspace*{0.2cm}
	\subfigure[ ]{%
		\includegraphics[width=0.45\textwidth]{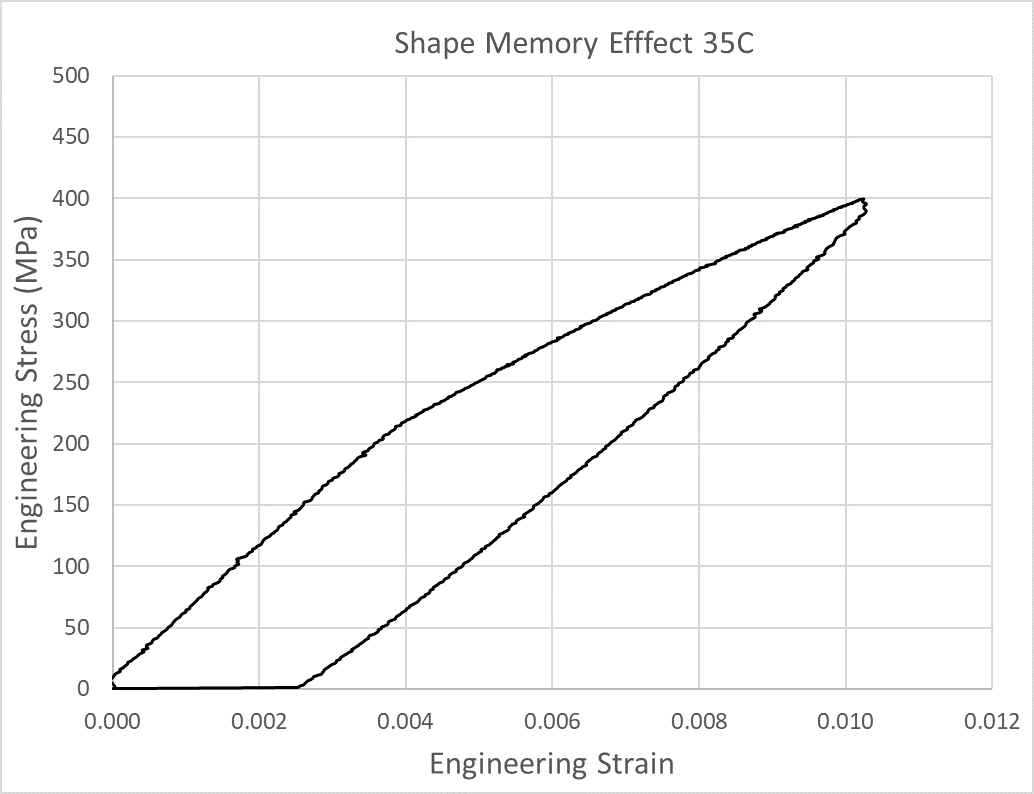}}%
	\caption{Uniaxial testing results (a) Pseudoelastic response in tension and compression. (b) Shape memory effect in tension. } \label{fig:uniaxial}
\end{figure}

\begin{figure}[ht]
	\centering
	\includegraphics[width=0.7\textwidth]{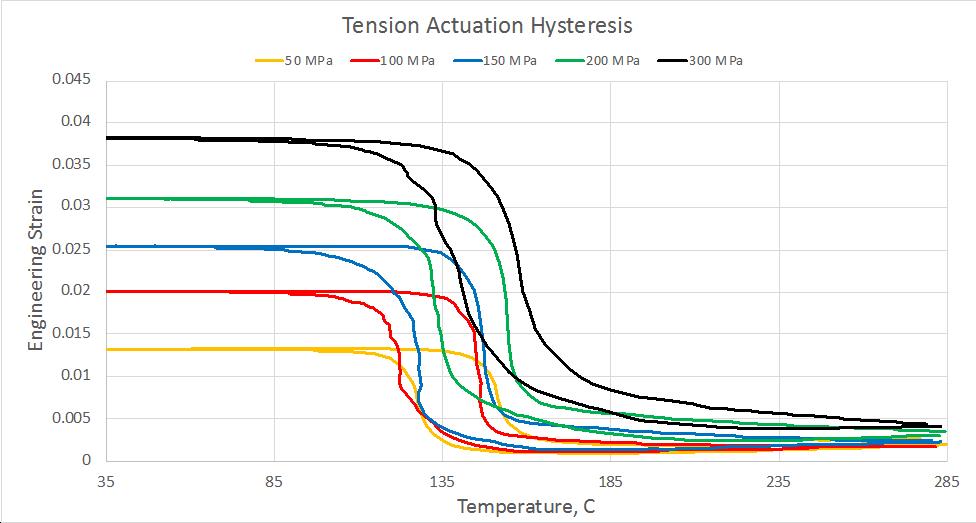}
	\caption{Actuation hysteresis for each stress level in the first cycle.}
	\label{fig:ACT_response}
\end{figure}
\section{Experimental Results}
\subsection{Differential Scanning Calorimetry}
The first step of the characterization process is to perform the DSC experiment on a small sample of the tested HTSMA materials. The latent heat release from the exothermic and endothermic phase transformation is captured to determine the phase transformation temperatures. Four DSC tests are performed on the HTSMA material with four different heat treatment conditions, i.e., as received, solutionized, solutionized and heat treated, and only heat treated conditions. The samples are vacuum sealed and water quenched if necessary for the treatment process. A selection is chosen based on stability and favorable transformation temperatures. Since the rod is hot extruded before being received, the best preparation option is the as-received condition with a 550 $^{\circ}$C/3hrs heat treatment. This gives transformation temperatures $M_f$ = 105 $^{\circ}$C, $M_s$ = 118 $^{\circ}$C, $A_s$ = 140 $^{\circ}$C, and $A_f$ = 155 $^{\circ}$C.

\begin{figure}[ht]
	\centering
	\includegraphics[width=0.7\textwidth]{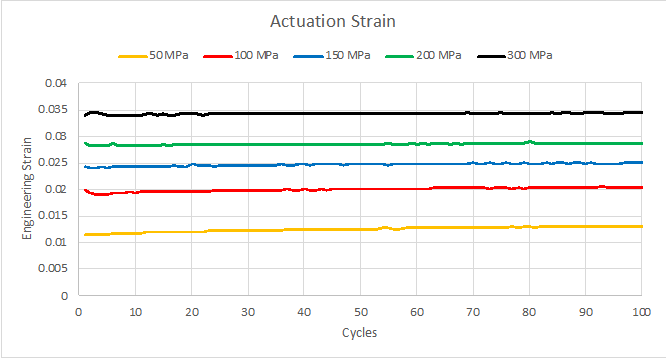}
	\caption{Actuation strain evolution for each stress level.}
	\label{fig:ACT_strain}
\end{figure}

\begin{figure}[ht]
	\centering
	\includegraphics[width=0.7\textwidth]{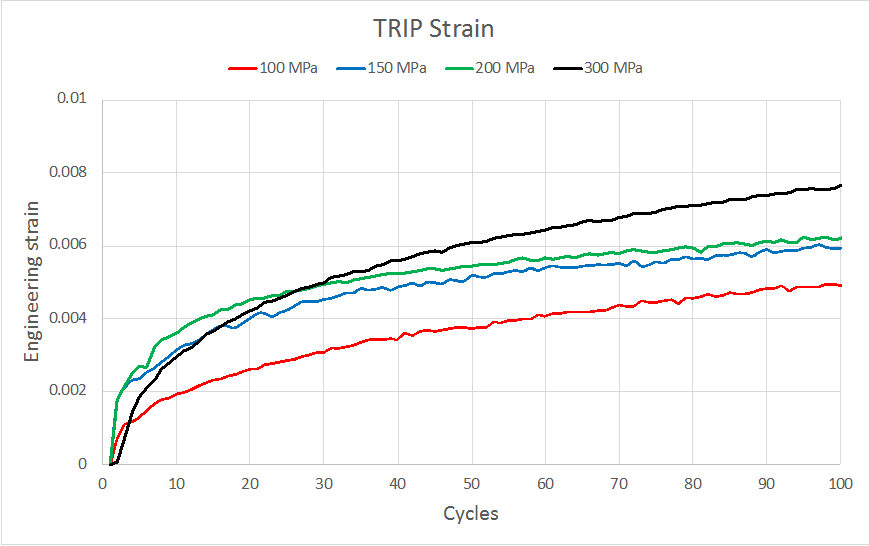}
	\caption{TRIP strain evolution for different stress levels.}
	\label{fig:TRIP_strain}
\end{figure}

\subsection{Pseudoelastic and Shape Memory Effect Response}
Pseudoelastic tests are also performed in order to get the basic SMA material parameters. These tests are conducted on the MTS insight load frame. The test is performed at a chamber stabilized temperature of 185$^{\circ}$C, 30$^{\circ}$C above $A_f$, and loaded from 0 to 11,400 N in crosshead control at a rate of 0.1mm/min as quasi-static loading conditions. This material is not expected to provide a pseudoelastic response with the questionable purity, but 4.6\% strain (out of the total 4.8\%) is recovered upon unloading and the material clearly exhibits the pseudoelastic effect. The same test procedure is performed in compression as well. The stress-induced transformation starts at 560MPa for the tension case, and 680MPa for the compression case. Compression results show that there is no flat stress plateau and the martensite reorientation requires additional nominal stresses to complete detwinning.  

A tension test to calculate the elastic modulus in the martensite phase is conducted as well, see Figure \ref{fig:uniaxial}. The Shape Memory Effect occurs when the deformation caused in martensite is recovered after heating into the austenite phase in a stress-free condition. The test is performed at chamber stabilized 35$^{\circ}$C, loaded from 0 to 4400 N in crosshead control at a rate of 0.1mm/min for quasi static conditions. The test results in 1\% strain at an average stress of 400 MPa. The strain recovered from heating the tested specimen in a stress-free condition was 0.25\%.

\subsection{Uniaxial Tension Actuation Cycling}
Constant load actuation tests were conducted at average stress values of 50, 100, 150, 200, and 300 MPa in order to characterize the thermomechanical response of HTSMA. The joule heating is applied and controlled between a 35$^{\circ}$C lower cycle temperature and a 285$^{\circ}$C upper cycle temperature, which produces full transformation cycles. Each test was performed on a different specimen right after processing. The specimens were cycled enough for characterization and quantification of plastic strain. The actuation hysteresis for each applied stress for the first cycle is shown in Figure \ref{fig:ACT_response}. The transformation temperatures and the actuation strain increase with increasing applied load. Each set of transformation temperatures for each applied stress level is calculated using the method of tangents in order to create the phase diagram in tension\cite{lagoudas2008}.

\begin{figure}[ht!]
	\centering
	\includegraphics[width=0.7\textwidth]{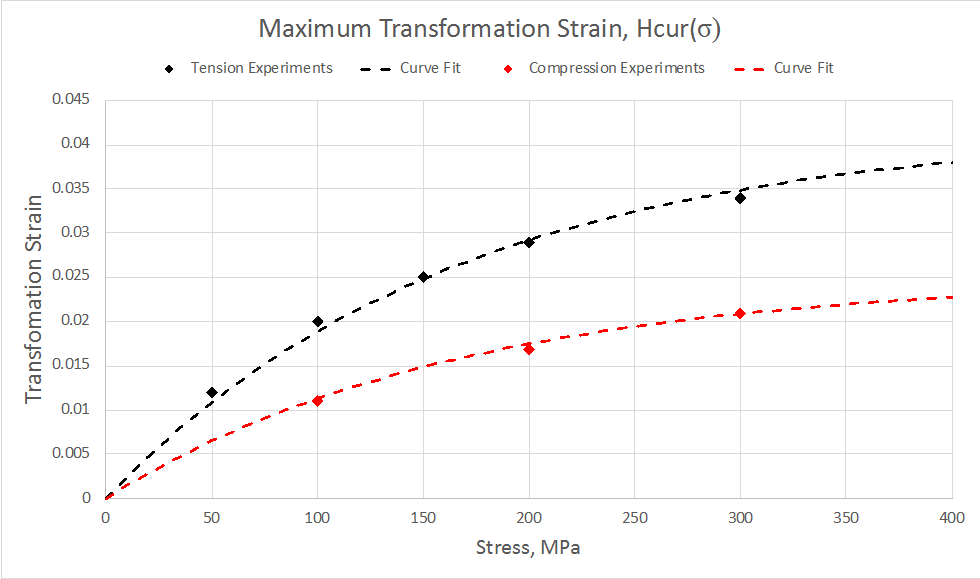}
	\caption{Maximum transformation strain from uniaxial tension tests at different stress levels.}
	\label{fig1:Hcur}
\end{figure}

\begin{figure}[ht!]
	\centering
	\includegraphics[width=0.6\textwidth]{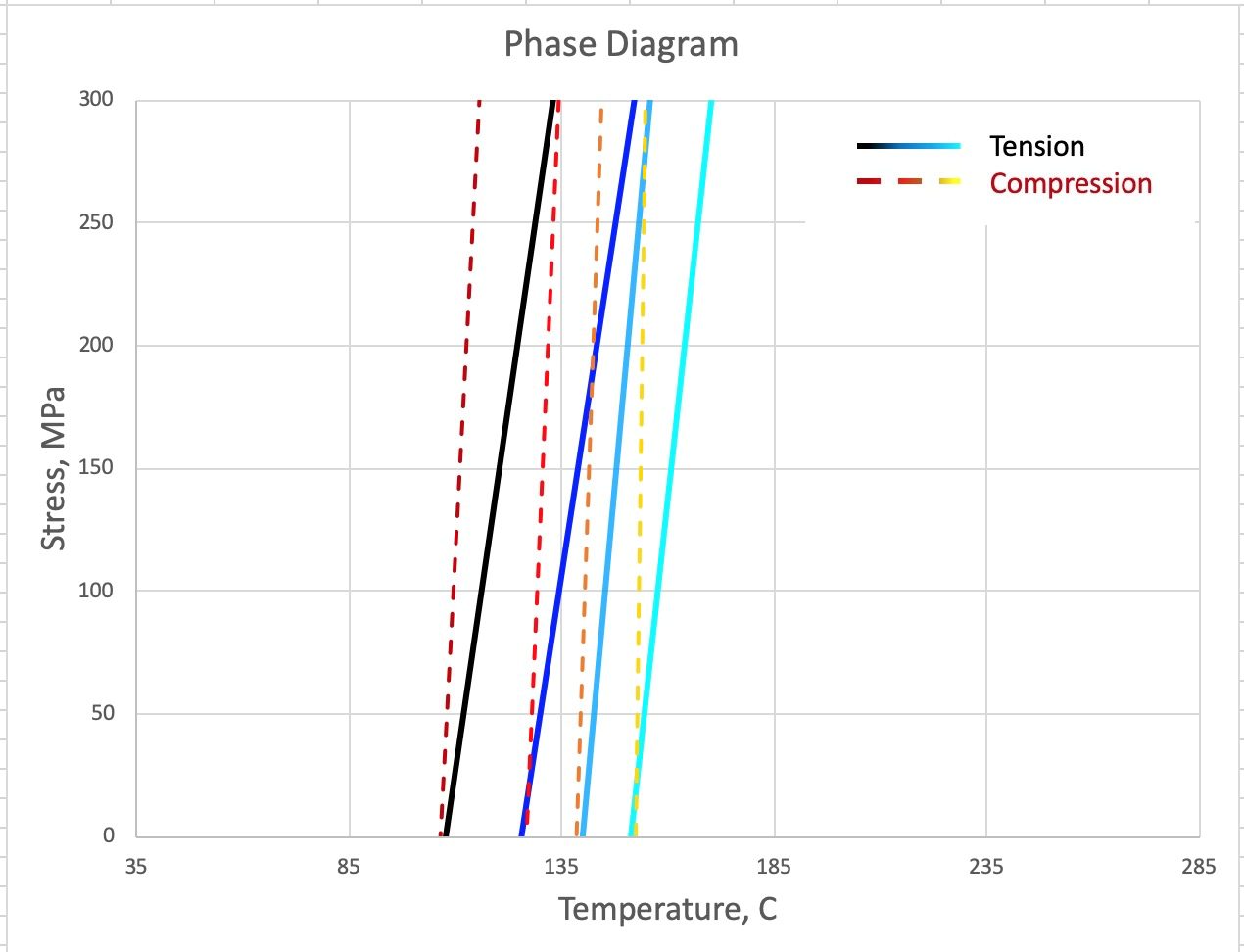}
	\caption{Phase diagram of the first actuation cycle in tension and compression.}
	\label{fig1:PhaseD}
\end{figure}

As shown in Figure \ref{fig:ACT_strain}, for each stress level, the total actuation strain remains constant throughout the actuation cycling and shows stability at least during the early stage in the fatigue life. Previous studies show that the actuation strain gradually decreases during the fatigue life due to softening and damage, but for the early fatigue life, it can be assumed constant \cite{wheeler2016actuator,calhoun2015actuation,phillips2019evolution}. Furthermore, plastic strains develop after each subsequent cycle as shown in Figure \ref{fig:TRIP_strain}. The TRIP strain development shows an expected trend with the applied stress levels, where increasing the applied load increases TRIP strain developed during cycling. The 50 MPa test is not shown here due to data acquisition errors.

\begin{figure}[ht!]%
	\centering
	\subfigure[]{%
		\includegraphics[width=0.43\textwidth]{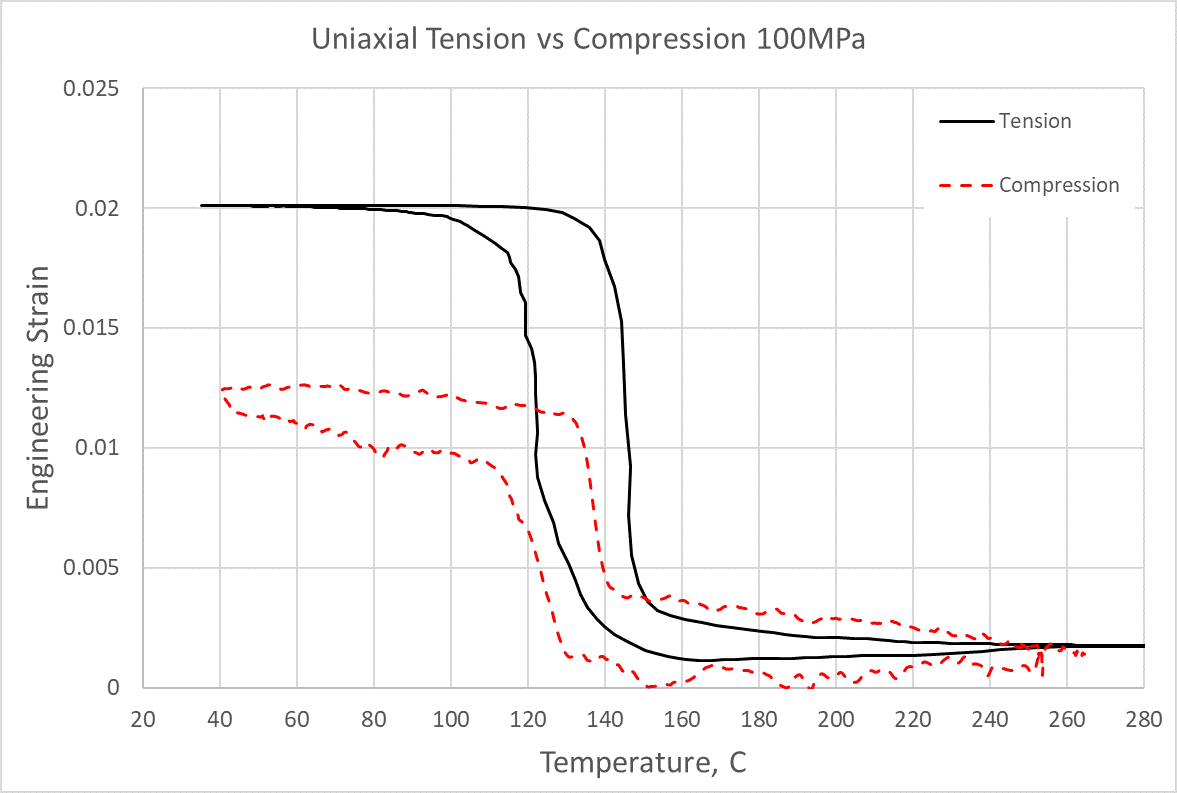}}%
	\hspace*{0.3cm}
	\subfigure[ ]{%
		\includegraphics[width=0.505\textwidth]{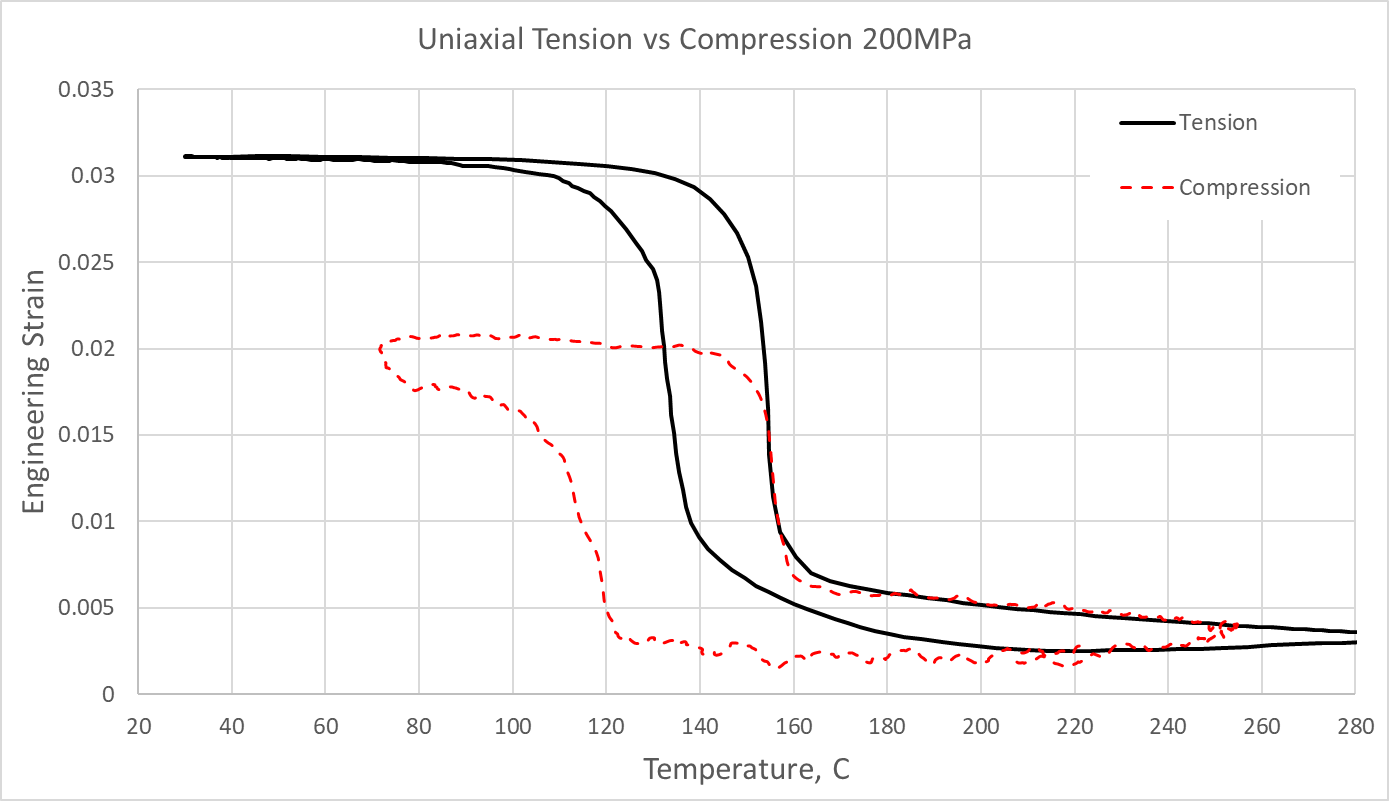}}%
	
	\centering
	\subfigure[]{
		\includegraphics[width=0.5\textwidth]{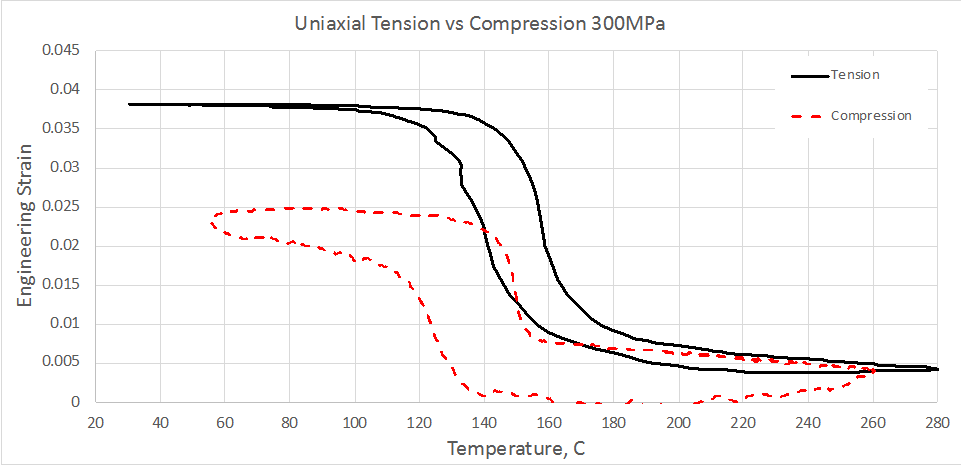}}	
	\caption{Tension-compression asymmetric actuation responses at different stress levels (a) 100MPa (b) 200 MPa (c) 300 MPa.}\label{fig:TC}%
\end{figure} 

\begin{figure}[ht!]
	\centering
	\includegraphics[width=0.8\textwidth]{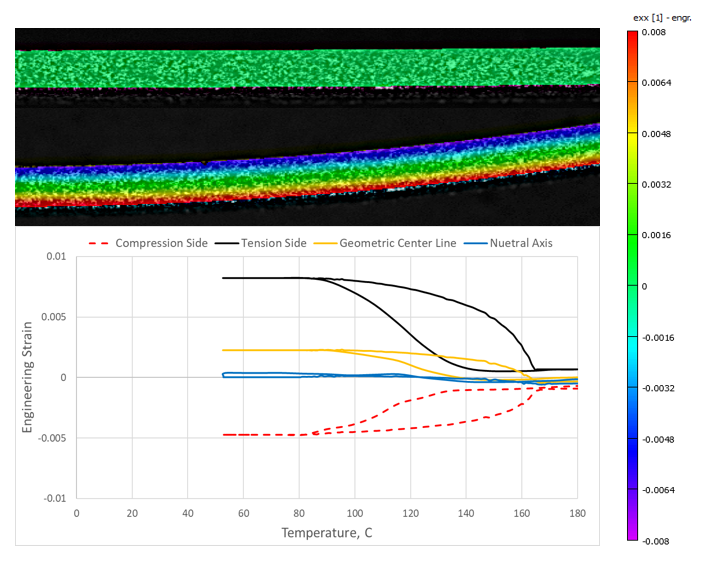}
	\caption{Localized engineering strain in the axial direction vs. temperature at the compression side, tension side, geometric center, and stress-free neutral axis during the first actuation cycle.}
	\label{fig:BeamResults}
\end{figure}

\subsection{Uniaxial Compression Actuation Cycling}
Three uniaxial actuation tests are performed in compression at the following nominal stress levels, i.e., 100, 200, 300 MPa. The results are compared against the tension cases at the first cycle is shown in Figure \ref{fig:TC}. In general, the maximum transformation strain is significantly lower in compression, and the transformation temperatures are also lower. High and low temperature regions where the thermal oven caused significant noise distortion for the DIC images are removed for clarity, but the data is sufficient for conclusions to be made and for modeling efforts. The transformation strain for each stress is quantified in Figure \ref{fig1:Hcur}. There is about a 63\% reduction in the transformation strain for the same applied stress in compression compared to tension. Measurements show the temperature on the specimen compared to the fixture temperature differed by around 9$^{\circ}$C during cooling and heating, meaning there is a slight temperature gradient across the sample. The results shown in compression are influenced by the temperature gradient slightly, broadening the hysteresis and the transformation temperature domain. The normalized compression data has been superimposed with the tension data to compare directly.

\begin{figure}[ht!]
	\centering
	\hspace{-1cm}
	\includegraphics[width=0.8\textwidth]{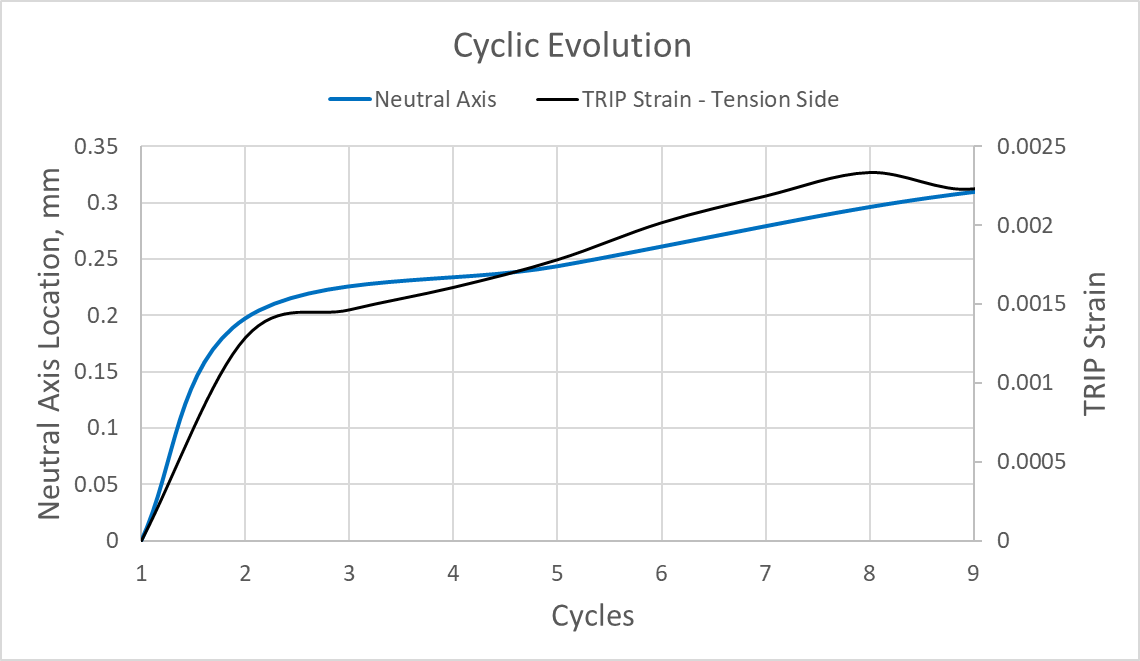}
	\caption{ Superimposed plot of neutral axis shifting distance and generated TRIP strain at the tensile side of the four-point beaning beam within 9 actuation cycles.}
	\label{fig:TRIP_neutral}
\end{figure}

\subsection{Four-Point Bending Actuation Cycling }
\subsubsection{Experimental Result }
The four point bending experiment consists of a constant 15.2N load, corresponding to a 40MPa flexural stress, and thermomechanically induced actuation cycling. Results show a significant displacement, up to 11mm at the midpoint, a maximum tensile strain of 0.8\%, and a maximum compressive strain of 0.48\% during forward phase transformation. The strain field at the midpoint section shows obvious strain asymmetry across the thickness as shown in Figure \ref{fig:BeamResults}. Continued cycling results in an increase in vertical displacement after each subsequent cycle, as there is unrecovered deformation after each cycle. As it can be expected that there is a twinned martensite core and a fully detwinned martensite region across the thickness of the beam. The twinned martensite core only has elastic and thermal strain contributions, while the detwinned region has strain contributions from thermomechanically induced phase transformation as well. In order to show the tension-compression asymmetry across the thickness, four strain points are extracted from the DIC image at the midpoint where the maximum displacement is recorded. The neutral axis location, where no axial strain is measured, is 0.2 mm above the geometric center during forward phase transformation, the response curve for those four point are plotted in Figure \ref{fig:BeamResults}.

In addition, as the cycling continued, the neutral axis kept on shifting into the compression side at a similar trend as the TRIP strain as shown in Figure \ref{fig:TRIP_neutral}. The relationship between the neutral axis shift and the plastic strain can be further understood by the summation of elastic, thermal, transformation, and plastic strains after a temperature cycle, where the only changes in strain is from plastic strain. A clear relationship is shown by Figure \ref{fig:TRIP_neutral} where the transformation induced plastic strain at the bottom of the plate and neutral axis differ only by a scalar.

\begin{figure}[ht!]
	\centering
	\includegraphics[width=0.7\textwidth]{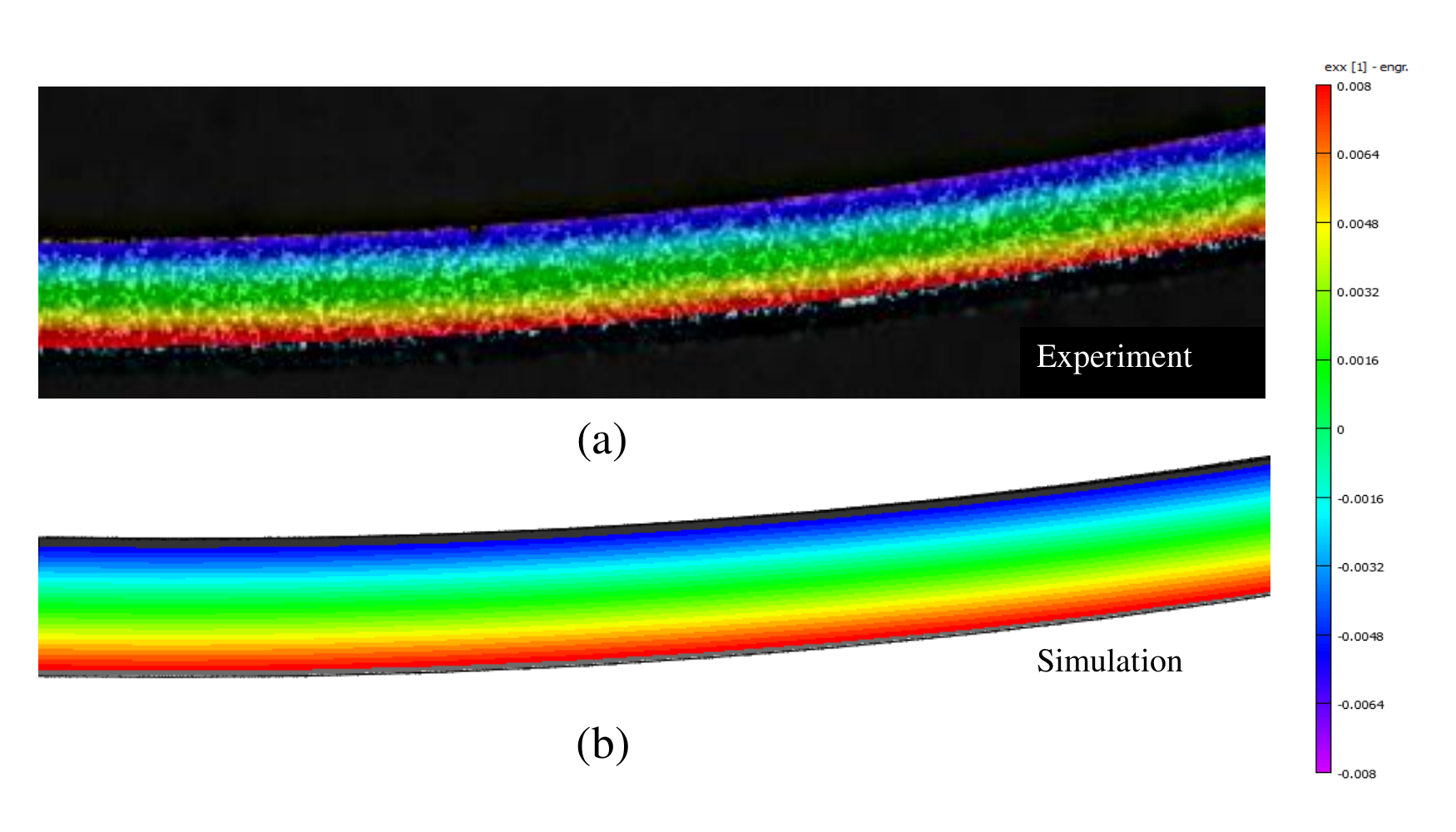}
	\caption{Axial engineering strain contour for the beam bending (a) Experiment and (b) Simulation.}
	\label{fig:Beam_contour}
\end{figure}

\begin{figure}[ht!]
	\centering
	\hspace{-0.8cm}
	\includegraphics[width=0.7\textwidth]{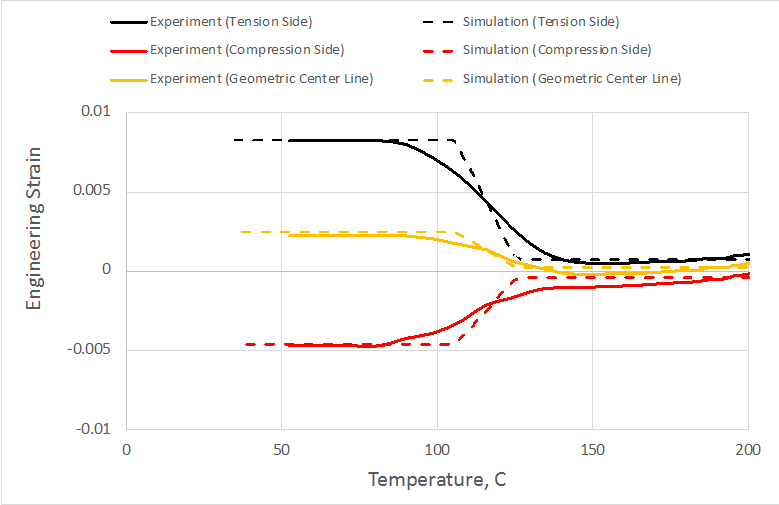}
	\caption{Comparison between experiment and simulation showing asymmetry in the bending response and a neutral axis shift at the geometric center during forward Phase transformation.}
	\label{fig:beam_simualtion}
\end{figure}

\subsubsection{Simulation Result }

\begin{figure}[ht!]
	\centering
	\hspace{-0.8cm}
	\includegraphics[width=0.7\textwidth]{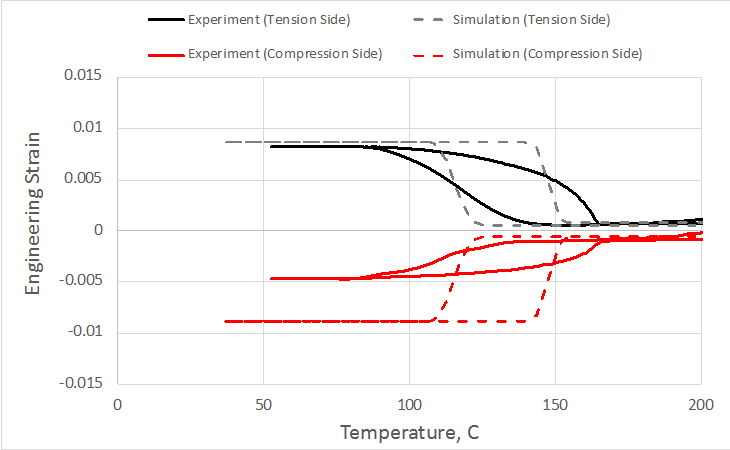}
	\caption{Simulation and experimental results comparison of four-point beam bending showing the engineering strain at the top (Compression side) and bottom (Tension Side) at the midpoint.}
	\label{fig:TC_simulation}
\end{figure}

\begin{figure}[ht!]
	\centering
	\hspace{-0.8cm}
	\includegraphics[width=0.7\textwidth]{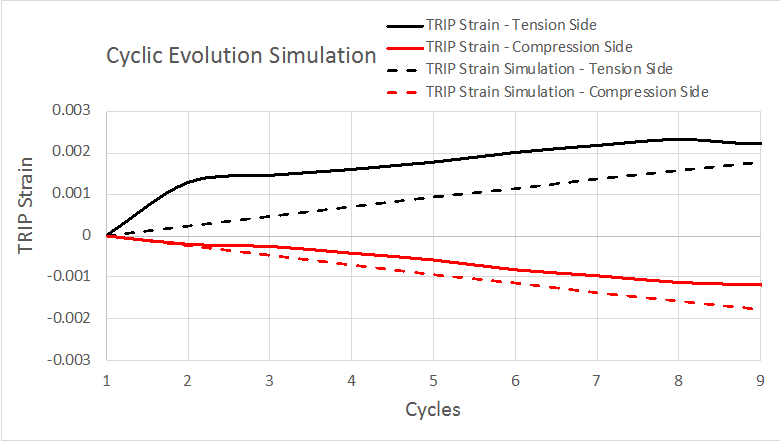}
	\caption{Generated TRIP strain comparison between simulation and experiment without tension-compression asymmetry during the actuation cycling of four-point beam bending test.}
	\label{fig:TRIP_Simulation}
\end{figure}

There has been a lot of constitutive modeling efforts been made during recent decades to predict the intrinsic material response of SMAs, a through review can be found from elsewhere\cite{xu2017AIAA,xu2018ASME,xu2019SMS,lagoudas2012}. As the motivation of this research is to facilitate the development of a phenomenological model that is able to capture the tension-compression asymmetry and TRIP during cyclic thermo-mechanical actuation loading, the results obtained from previously performed tension/compression and bending experiments are used for the formulation and validation of the SMA constitutive model capturing the anisotropic phase transformation characteristics. In order to achieve the desired modeling capability, modifications on the transformation functions are needed. In general, there are various formulations to predict the tension-compression asymmetry feature based on different transformation functions, i.e., the $J_2$, the $J_2-I_1$ and the generalized $J_2-J_3-I_1$ theory based approaches. Specifically, $J_2$ based models cannot capture tension–compression asymmetry; $J_2-I_1$ based models can capture tension-compression asymmetry or pressure dependence, and $J_2-J_3-I_1$ based models capture volumetric transformation strain, asymmetry, and pressure dependence \cite{boyd1996thermodynamical,hartl2018computationally}. Due to the inherent complexity of the model formulations, the authors advise the interested readers to reference the available publications \cite{qidwai2000,hartl2018computationally} for more details regarding the model derivations and implementations. In the following simulation work, a three-dimensional finite element analysis model was developed to match the beam bending boundary value problems. The finite element model consists of 1/4 th of the total structure to represent symmetry along the length and width to reduce computational effort. A combination constitutive model based on the work of Xu et al.\cite{xu2019AIAA} and Hartl et al.\cite{hartl2018computationally} is used in this simulation.

The strain contour in the longitudinal direction across the beam thickness is shown in Figure \ref{fig:Beam_contour}, it can be seen that the simulation matches the experiment displacement and strain relatively closely by considering the tension-compression feature through the thermomechanical cycling. In addition, as shown in Figure \ref{fig:beam_simualtion}, the comparison between experiment and simulation shows there is a deformation asymmetry in the bending response and a neutral axis shift at the geometric center during the forward Phase transformation. However, it is should be noted that the maximum compression strain can be over predicted without considering tension-compression asymmetry, see Figure \ref{fig:TC_simulation}. Moreover, the plastic strain developed can be under predicted in the tension side and over predicted in the compression side without considering the tension-compression asymmetry feature. In summary, there is good correlation between the experimental results and the simulation results using the $J_2-J_3$ transformation surface, and the simulation qualitatively captures the macroscopic neutral axis shift by plotting the strain at the geometric center, the model is also able to predict the generated TRIP strain under actuation bending cycling, but it is still in need of further improvement.

\section{CONCLUSION}
In this work, tension, compression, and bending actuation cycling tests on HTSMA specimens are performed by the custom-built test frames to facilitate the development and validation of newly proposed SMA constitutive model considering the cyclic anisotropic phase transformation characteristics for HTSMAs. The performed experiment shows an intrinsic phenomenon for HTSMAs under cyclic thermomechanical conditions, i.e., TRIP strain accumulation and tension-compression asymmetric phase transformation. During the phase transformation in an SMA beam under four-point bending, tension-compression asymmetry occurs and the zero-strain neutral axis shifts as a result of the asymmetric generation of TRIP strains on different sides of the beam. The combination of tension-compression asymmetry and the formation of TRIP strains during thermomechanical actuation cycling is measured and simulated by using a recently developed phenomenological model, and there is good correlation between the experimental results and the simulation, but the model is still in need of further improvement. 

Future work along this paper can mainly focus on the development and improvement of the current SMA constitutive model considering tension-compression asymmetry, TRIP strain, thermo-mechanically coupling, and partial phase transformation loops through an unified large deformation framework. Additional experiments could be conducted to validate the proposed modeling capabilities for additional features. Also, experimental results from torque tubes under actuation cycling loading would provide more model validation data and component performance for HTSMAs actuators. Lastly, testing HTSMA specimens with notches, or localized stress concentrations and using the constitutive model to predict the cyclic response would be revolutionary for the analysis of HTSMA-based actuators.

\section{Acknowledgments}\label{Ack}
This work is supported by the National Aeronautics and Space Administration (NASA) through the University Leadership Initiative (ULI) project under the grant number: NNX17AJ96A. The conclusions in this work are solely made by the authors and do not necessarily represent the
perspectives of NASA.

\bibliographystyle{aiaa} 
\bibliography{myarticle}

\end{document}